\documentclass[aps,prl,twocolumn,superscriptaddress,nofootinbib,
nobalancelastpage,showpacs,nobibnotes]{revtex4}
\usepackage{epsfig}

\begin{document}
\title
{Polarization exchange in colliding photon beams in a medium of an atomic gas.}

\author{R. F. Sawyer}
\affiliation{Department of Physics, University of California at
Santa Barbara, Santa Barbara, California 93106}

\begin{abstract}
Photon-photon interactions mediated by an atomic gas can effect efficient polarization exchanges between two beams, leaving the medium
exactly in its initial state. In, e.g., hydrogen, the distance required for macroscopic exchange is of order one tenth the distance in which
the ordinary non-linear index of refraction would induce a phase change of pi. Several examples are worked out that show the variety
of behaviors that can result, depending on the initial respective polarizations stated and the angle between the beams. Of particular interest are initial 
conditions
in which there is no exchange at a mean field level, conventionally believed to apply when the number of photons, N, is large. Then the full theory
leads both to large exchange and to large entanglement between the beams.
Our most solid results indicate that one would have to wait a time proportional to log[N] to see this effect, but there are some indications that this
behavior can be circumvented.

\end{abstract}
\maketitle 
\section{1. Introduction}
There is an extensive literature on the theoretical and experimental aspects of creation of entanglement in systems
in which a number of spins interact with themselves and with external fields or probes. 
The work reported here is rather parallel in some ways, but it is concerned with photon polarizations rather than
particle spins.
The ideal would be the creation of two separated pulses of photons, A and B, with the polarization state of the
whole system being a quantum superposition $|\,A_1\rangle\,|\,B_2\rangle+|\,A_2\rangle\,|\,B_1\rangle$, ``cats" in current parlance, where, for example, for co-linear beams the subscripts, 1,2 indicate, respectively, one direction of plane polarization and the perpendicular one.

In our models the creation of coherent mixtures from two initially completely polarized beams is accomplished by their mutual interaction mediated by a gaseous medium, the state of the medium remaining unchanged. No outside intervention is required, except for the introduction of the
beams into the reaction region.
The dynamics of the polarization evolution is determined by an effective Hamiltonian, $H_{\rm eff}$, that operates purely within the polarization space for the
combined beams.

In our concrete modeling we choose atomic hydrogen as the medium, impractical as it would be for implementation, for the reason that
it is in hydrogen that we can do a definitive calculation of the effective photon-photon interaction. The intent of the paper in any case is to give schematic
arguments, with the calculation in hydrogen serving to provide only rough orders of magnitudes.

There are several features of our system and the results that emerge that are particularly distinctive:
 
 1).The entanglement that is produced is between the polarizations of two photon beams during interactions that do not change
 the momentum or direction of any photon. The beams enter an interaction region from different directions, and then both leave
 the region.  They can emerge as entangled ``cats" and remain that way until they encounter some polarization-dependent interaction with an environment.

2). There are many possible variations of initial polarizations, final detector parameters, and angle between the beams.
Numerical experiments on some miscellaneous choices give a variety of interesting ways in which the beams can develop macroscopic entanglement of one
beam with the other.

3). There is a general question that arises with respect to our results; namely, is it a theorem that to have two systems, each of $N $ ``spins", to become macroscopically entangled from their mutual interaction in a medium, requires a time proportional to $[\log N]$ for large $N$, (taking fixed number densities $N$/Vol.)? We might infer this from our
simplest, and probably most reliable, examples. Indeed, similar behavior is found in other quantum systems, e.g., in
models of a two-mode boson condensates \cite{anglin}-\cite{anglin2}. But we have some hints that the $[\log N]^{-1}$ slowing may be avoidable 
under certain conditions.

4). In a related matter, when the initial state can be characterized at a mean field level as one of unstable polarization equilibrium, then
one can roughly define a ``quantum break time", since for large $N$ the system remains quiescent for a period proportional to $[\log N]$, and then
flips completely to another state over a short transition time that is independent of $N$.  Without recourse to approximate calculations in specific models can one prove a linear dependence of this ``quantum-break time" on the logarithm of the number of degrees of freedom?

We emphasize that the rates that we calculate have almost nothing to do with photon-photon cross-sections in media (as discussed, e.g. , in ref. \cite{king}).
Photon-photon cross-sections are very, very small. Indeed, long ago Kotkin and Serbo \cite{ks} (also see \cite{rfs1}) showed how, for the vacuum case governed by the Heisenberg-Euler effective interaction \cite{iz}
polarization exchange occurs in a distance many orders of magnitude smaller than [cross-section$\times$number density]$^{-1}$.

\section{2. Effective coupling in an atomic medium.}
We consider the reaction $\gamma ({\bf q_1})+\gamma ({\bf q_2})\rightarrow \gamma' ({\bf q_1})+\gamma' ({\bf q_2})$, where the momenta of the photons are unchanged, where the primes on the right signify that the polarizations
have changed, and where all four photon interactions are with a single atom. 
For concreteness we take $\bf q_1$ to be in the $\hat{\bf z}$ direction with corresponding polarization basis vectors, $\vec \xi_{1}=\hat{ \bf x}$,
 $\vec \xi_{2}=\hat{\bf y}$; and  $\bf q_2$ in the x-z plane at an angle $\pi-\theta$ to the z axis, with polarization vectors $\vec \eta_{1}=\hat{\bf x}  \cos \theta +\hat{\bf z}\sin \theta$, 
$\vec \eta_{2}= \hat{\bf y}$ .

We calculate our effective Hamiltonian in this real basis of plane polarization states. We define $H_0$ as the Hamiltonian for the atom plus the free electromagnetic field, and
$H_e$ as the non-relativistic electron-photon coupling $H_e =(2 m_e)^{-1} (-e\,\vec p\cdot \vec A-e\,\vec A\cdot \vec p+ 
e^2 \vec A \cdot \vec A)$.  We write out the textbook expression for $\vec A$ , in order to define notation and to clarify factors of volume, $V$,  in what follows, including only the terms that refer to the four modes enumerated above,
\begin{eqnarray}
&\vec A=(2\omega_1 V)^{-1/2} (a_1 \vec \xi_{1}+a_2 \vec \xi_{2} )e^{-i \omega_1(t-z)}+
\nonumber\\
&(2 \omega_2 V)^{-1/2}(b_1\vec \eta_{1}+b_2 \vec \eta_{2} )e^{-i \omega_2 t +i \bf q_2  \cdot r}+H.C.\,,
\end{eqnarray}
where the operators $a_1,a_2$ annihilate photons in the $ \bf q_1$ direction, with the respective polarization
vectors $\zeta_{1a},\zeta_{1b}$; with the operators $b_1,b_2$ annihilating photons in the $ \bf q_1$ direction with polarizations $\zeta_{2a},\zeta_{2b}$. 
We define
\begin{eqnarray}
T_\gamma=\langle {0}, {\vec  p}_{\rm at}|\,H_e [1-{1\over E_0- H_0}H_e]^{-1}\, |   { 0}, {\vec  p}_{\rm at}\rangle\,,
\end{eqnarray}
which is the conventional  ``T matrix" operator, where we have taken an expectation in a ground state, $|{\rm 0} ,{\vec p}_{\rm at} \rangle$, in the space of the atom,
and no momentum transfer to the atom as a whole. Here $T_\gamma$ is still an operator in the photon space, a function of the four annihilation operators $a$ and $b$ and their associated creation operators.

Next we extract the part of $T_\gamma$ that is fourth order in the coupling.  Designating this piece 
as $H_I^{\rm eff}$, we have
\begin{eqnarray}
&H_I^{\rm eff}=
\langle {0}, \vec p_{\rm  at}|
\Bigr (H_\gamma {1\over E_0- H_0}H_ \gamma {1\over E_0- H_0}H_ \gamma 
{1\over E_0- H_0}H_ \gamma\Bigr) 
| {0} , \vec p_{ \rm at}\rangle
 \nonumber\\
& + \vec A\cdot \vec A~{\rm  terms}~,~~~~~~~~~~~~~~~~~~~~~~~~~~
\label{h23}
\end{eqnarray}
where $H_\gamma\equiv(2 m_e)^{-1} (-e\,\vec p\cdot \vec A-e\,\vec A\cdot \vec p)$, the single photon emission vertex. (The double photon vertices
from the $\vec A\cdot \vec A$ terms have also been included in all calculations in this paper.)
The part that describes our reaction will have
a sum of terms with various factors like $a_1^\dagger a_1 b_1^\dagger b_1$, $a_1^\dagger a_2 b_2^\dagger b_1$, etc. It is convenient in what follows to replace these operators with ones constructed
 from,
 \begin{eqnarray}
 s_+=a_2^\dagger a_1~~,~~  s_-=s_+^\dagger~~,~~ t_+=b_2^\dagger b_1~~~,~~~
  t_-=t_+^\dagger\,,
   \nonumber\\
 ~
 \nonumber\\
 s_3=a_2^\dagger a_2-a_1^\dagger a_1  ~~~~,~~~~ t_3=b_2^\dagger b_2-b_1^\dagger b_1 \,,
 \nonumber\\
  ~
 \nonumber\\
 s_0=a_1^\dagger a_1+a_2^\dagger a_2  ~~~~,~~~~ t_0=b_1^\dagger b_1+b_2^\dagger b_2\,.
 \label{ops}
 \end{eqnarray}

The connection between the characterization of polarizations as used in this paper and the Stokes parameters, Q,U,V, 
\cite{mcm}
is provided by writing the polarization density matrix for, e. g., our beam \# 1 as, 
\begin{eqnarray}
\rho=N^{-1}\sum_i[ s_0^i+s_3^i  Q+(s_+^i+s_-^i)U+i(s_+^i-s_-^i)V]\,,
\label{stokes}
\end{eqnarray}
where $s^i$ is the operator for the $i$'th photon in the beam.

We work in the dipole approximation for the interaction of the photons with the atomic electrons. Then the only place that
the angle between the two beams enters will be in the polarization vectors themselves.
There are a plethora of terms that result from attaching the four photon lines of our process to the atomic electron
line in all possible ways. Fig. 1 gives a graphical representation of one of these ways.
There is further proliferation when we assign the values photon polarizations in all possible ways. In this diagram are shown polarization vectors, $\vec \xi,\vec \xi'$
for the initial and final states of the $\vec q_1$ photons, and drawn in any way from the basis set $\vec \xi_1,\vec \xi_2$ introduced above, and similarly
with the vectors $\vec \eta$, $\vec \eta'$ for the $\vec q_2$ photons. 
\vspace{-1 in}

\begin{figure}[h] 
 \centering
\includegraphics[width=4 in]{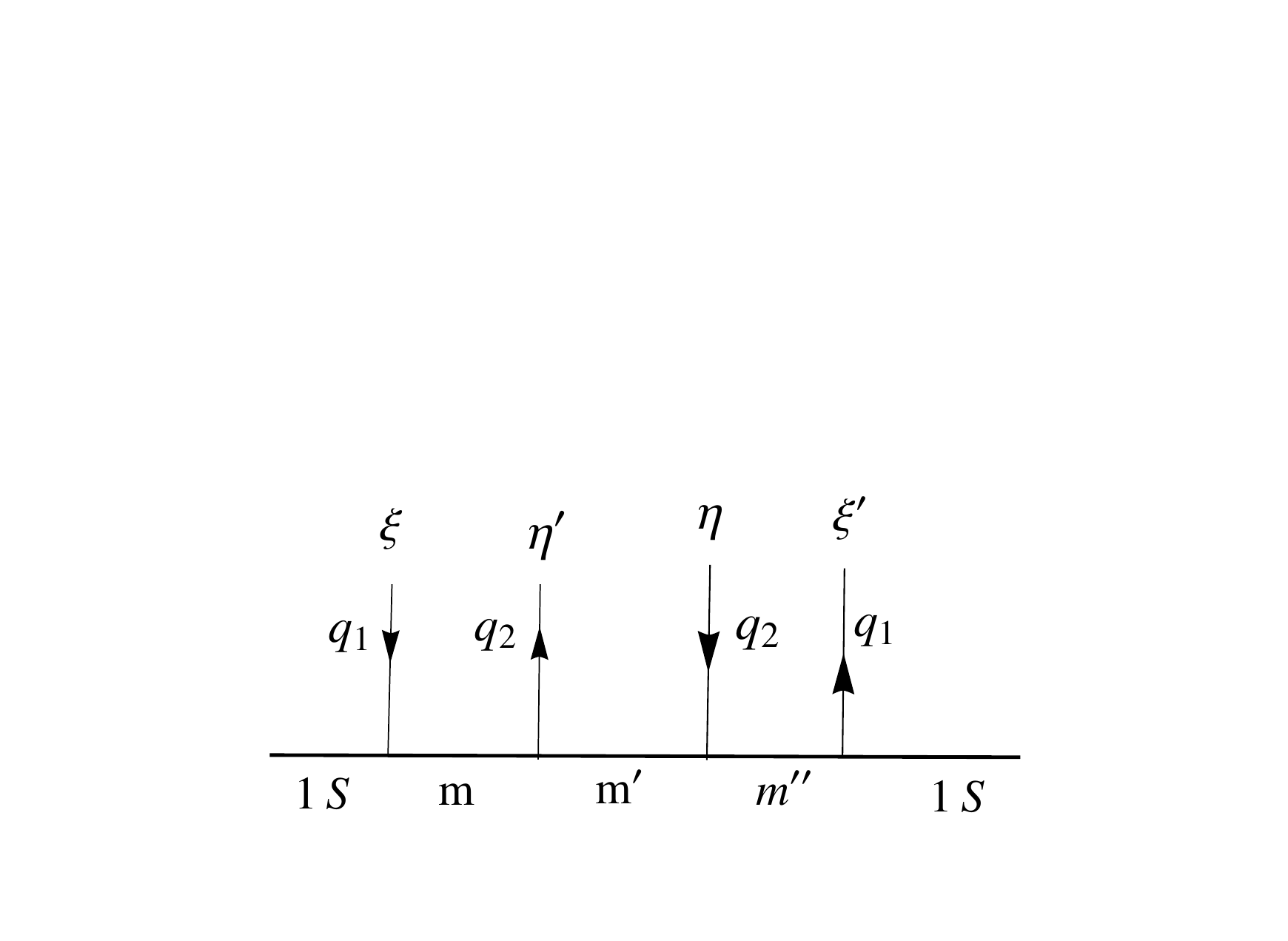}
 \caption{ \small An example of a graph contributing to $H_{\rm eff}$ for polarization evolution. The heavy horizontal line stands for the atom.
     The labels $m,m',m''$ denote the excited levels of the atom. 
     The photons can be attached in 24 different orders}   
 \label{fig. 1}
\end{figure}

For example, in the dipole approximation the contribution from the single graph of fig. 1 is,
\begin{eqnarray}
&H_I^{\rm eff}={ \alpha^2 \omega_1^2 \omega_2^2\over  4 \omega_1 \omega_2  V^2  }
\nonumber\\
&\times\sum_{m,m',m''}{\langle 0|\vec r\cdot  \vec \xi^{\,\,'} |m''\rangle\langle m''|\vec r\cdot  \vec \eta\, |m'\rangle  
\langle m' |\vec r\cdot  \vec \eta^{\,\,'} | m\rangle \langle m|\vec r\cdot  \vec \xi\, |0\rangle 
\over (\epsilon_0+\omega_1-\epsilon_{m''})(\epsilon_0+\omega_1-\omega_2-\epsilon_{m'})(\epsilon_0+\omega_1-\epsilon_{m})} ,
\nonumber\\
\,
\label{bigsum}
\end{eqnarray}
where now $|0\rangle$ denotes the atomic ground state, $| m\rangle$ an excited state and $\vec r$ is the position operator for the atomic electron.\footnote{ In (\ref{bigsum})
we have used the universally accepted lore that replaces the operator $\vec p \cdot \xi/m_e$
of the canonical formulation by $\omega \vec r \cdot \xi$, and at the same time discards the $\vec A\cdot \vec A$ term in $H_I$.
Rather quixotically, not completely sure that this was correct for the four-photons-attached case, we also calculated the result in hydrogen
directly from the canonical approach, obtaining the same answer as is calculated (much more simply) from (\ref{RH}).}

 The atomic motion label $\vec p_{\rm at}$ has been dropped, since in selecting the totally coherent part
 of the photon process we leave the atom exactly its original state of motion, and the thermal velocities of the atoms are
 small enough to make doppler shifts inconsequential as well. Eq. (\ref{bigsum}) is equivalent to general expressions
 that we find in non-linear optics books \cite{B}, though a certain amount of translation is required to get from ``generalized
 polarizabilities" to our effective Hamiltonian. Furthermore, we have not found the complete polarization-exchange terms exhibited explicitly in the literature.

The results of the fourth order perturbation calculation
at the dipole level are expressed in either case in terms of initial and final polarization vectors $\vec \xi,\vec \xi' $
for the respective initial and final polarizations of the $\vec q_1$ beam, and  $\vec \eta,\vec \eta' $ for those of the
$\vec q_2$ beam and are of the form,

 \begin{eqnarray}
& \langle \xi',\eta' |H_{\rm eff}|\xi,\eta \rangle={R  \over V}[\,( \vec  \xi \,' \cdot  \vec \eta \,' )( \vec  \xi \cdot  \vec \eta )+ 
 ( \vec  \xi  \cdot  \vec \eta \,' )( \vec  \xi\,' \cdot  \vec \eta ) \,]
 \nonumber\\
&+\lambda ( \vec  \xi  \cdot  \vec \xi \,' )( \vec  \eta\,' \cdot  \vec \eta ) \,] \,,
\label{ha}
 \end{eqnarray}
where $R$ is an intensive quantity that is proportional to the electron density $n_e$, and where we have gone from the
case of a single atom to that of a gas, taking advantage of the fact that zero momentum is transferred to the gas and trading one factor of the box volume inverse, $V^{-1}$, for a factor of $n_H$, the hydrogen density. This left one remaining
factor $V^{-1}$ from the two in (\ref{bigsum}). As a measure of strength we have calculated $R$ for atomic hydrogen
(not the most practical gas for implementation, but the one in which we can do an exact calculation in the dipole limit),
\begin{eqnarray}
R={ 2529 \pi^2 e^4 \omega^2 n_e  \over 8 m_e^2 {\rm [Ry]}^5}\,,
\label{RH}
\end{eqnarray}
where [Ry] is the Rydberg.
For the purposes of the present paper we can drop the $\lambda$ term in (\ref{ha}), which leaves the polarizations of the 
separate beams untouched. The angle $\theta$ between the two beams enters only implicitly here through
the orthogonality of the polarization space of each beam to its respective direction of propagation. The
translation of the matrix elements in (\ref{ha}) into those of operators constructed from $\vec s$ and $\vec t$, of
(\ref{ops}) is routine and gives,
\vspace{.4 in} 
\begin{eqnarray}
H_{\rm eff}= {R n_e \over  V}[-\sin^2 \theta\,\, s_3 t_0 -\sin^2 \theta\,\,s_0 t_3+
\nonumber\\
2 \cos\theta  \,     s_1 t_1+(1+\cos^2\theta)\,s_3 t_3]\,,
\label{ha2}
\end{eqnarray}
where $t_1=t_++t_-$, $t_2=(t_+-t_-)/i$, and similarily for $s_1, s_2$. 

It is a simple matter to transform to a circularly polarized basis. 
This is effected through $s_3\leftrightarrow-s_2$,  $t_3\leftrightarrow-t_2$, $s_1\leftrightarrow s_1$,$t_1\leftrightarrow t_1$,
\begin{eqnarray}
\hat H_{\rm eff}= {R \over V}[\sin^2 \theta\,\, s_2 t_0+\sin^2 \theta\,\, s_0 t_2+
\nonumber\\
2 \cos\theta  \,     s_1 t_1+(1+\cos^2\theta)\,s_2 t_2\,]
\label{ha2circ}
\end{eqnarray}
The eigenstate of $\sigma_3=1$ now is right circularly polarized and that of  $\sigma_3=-1$ is left-circularly polarized, with the conventions that we have adopted.

The interaction, (\ref{ha2circ}), is still between just two photons and dependent on the volume of the box. Now we extend to the case of two 
photon beams. The final equations of evolution for the polarization density matrix will not care whether all the photons in a single beam are in
a single coherent state or are distributed, with differing phases, over a narrow range of energies and directions. For our derivation
we choose the latter case.
  
In the ensemble of $N_1$ photons in one beam and $N_2$ in the other,  we designate
the constituent operators for the two beams by $s_\alpha^{(i)}$ , $i=1,...N_1$, and  $t_\alpha^{(j)}$,
$j=1,...N_2$, and define collective operators for the respective beams by $\sigma_\alpha=N_1^{-1}\sum_i [s_\alpha^{i}]$,
 $\tau_\alpha=N_1^{-1}\sum_i [t_\alpha^{i}]$.
The Hamiltonian for everybody interacting with everybody is then,
\begin{eqnarray}
&\hat H_{\rm eff}= {N_1 N_2 R\over V}[\sin^2 \theta\,\, \sigma_2 +\sin^2 \theta\,\, \tau_2+2 \cos\theta  \,     \sigma_1 \tau_1+
\nonumber\\
& (1+\cos^2\theta)\,\sigma_2 \tau_2\,]\,.
\label{ha3circ}
\end{eqnarray}
After rescaling time with the factor $R$
we use the Hamiltonian (\ref{ha3circ})  to obtain the Heisenberg equations, $i (d/dt) \sigma_\alpha=R_a^{-1}[ \sigma_\alpha, H^{\rm eff}]$ etc.. Then applying the commutation rules,
\begin{eqnarray}
[\sigma_\alpha,\sigma_\beta ]=2 N_1^{-1} i \epsilon _{\alpha\beta\gamma }\sigma_\gamma~, ~
[\tau_\alpha ,\tau_\beta]=2 N_2^{-1} i \epsilon _{\alpha \beta\gamma}\tau_\gamma,
\end{eqnarray}
and shifting to variables, $\sigma_\pm=(\sigma_1\pm i \sigma_2)/2$,
 $\tau_\pm=(\tau_1\pm i \tau_2)/2$, we obtain,
\begin{eqnarray}
i \dot \sigma_+=n_2  \sigma_3[ \sin^2 \theta+ \tau_+(1+\cos \theta)^2- \tau_-(1-\cos\theta)^2 ]\,,
\nonumber\\
\label{eom1}
\end{eqnarray}
\begin{eqnarray}
&i \dot \sigma_3= 2 n_1 [\sin^2\theta (\sigma_+ -\sigma_-)
-(\sigma_+\tau_+-\sigma_- \tau_-)(1-\cos \theta)^2
\nonumber\\
&+(\sigma_+\tau_- -\sigma_- \tau_+)Ê(1+\cos \theta )^2]\,,
\label{eom1a}
\end{eqnarray}
and,
\begin{eqnarray}
i \dot \tau_+=n_2  \tau_3[ \sin^2 \theta+ \sigma_+(1+\cos \theta)^2- \sigma_-(1-\cos\theta)^2 ]\,,
\nonumber\\
\label{eom2}
\end{eqnarray}
\begin{eqnarray}
&i \dot \tau_3= 2 n_1 [\sin^2\theta (\tau_+ -\tau_-)
-(\sigma_+\tau_+-\sigma_- \tau_-)(1-\cos \theta)^2
\nonumber\\
&-(\sigma_+\tau_- -\sigma_- \tau_+)Ê(1+\cos \theta )^2]\,.
\label{eom2a}
\end{eqnarray}

Here photon densities, $n_{1,2}=N_{1,2}/V$, have replaced photon number.
The equations for
$\tau_+$ and $\tau_3$, are obtained by $\vec \sigma\leftrightarrow \vec \tau$, $n_2\rightarrow n_1$. The set is closed when we add
$\sigma_-=\sigma_+^\dagger$, $\tau_-=\tau_+^\dagger$. Self-interactions within one beam do not appear in the above, since they produce no
polarization dependent terms in the equations.

The equations (\ref{eom1})-(\ref{eom2a}) are operator equations in an $N_1 N_2$ dimensional space of states. But we begin by addressing them in a
``mean field" approximation (MFT). In the present context this means that the four variables in the nonlinear equations (\ref{eom1}), (\ref{eom2}) are each replaced
by an expectation value, e.g.  $\sigma_+\rightarrow \langle \sigma_+\rangle =\langle \sigma_-\rangle ^* $. 
The resulting
equations are not exact, since the expectation of a product
is not the product of the individual expectations. 

\begin{figure}[h] 
 \centering
\includegraphics[width=3 in]{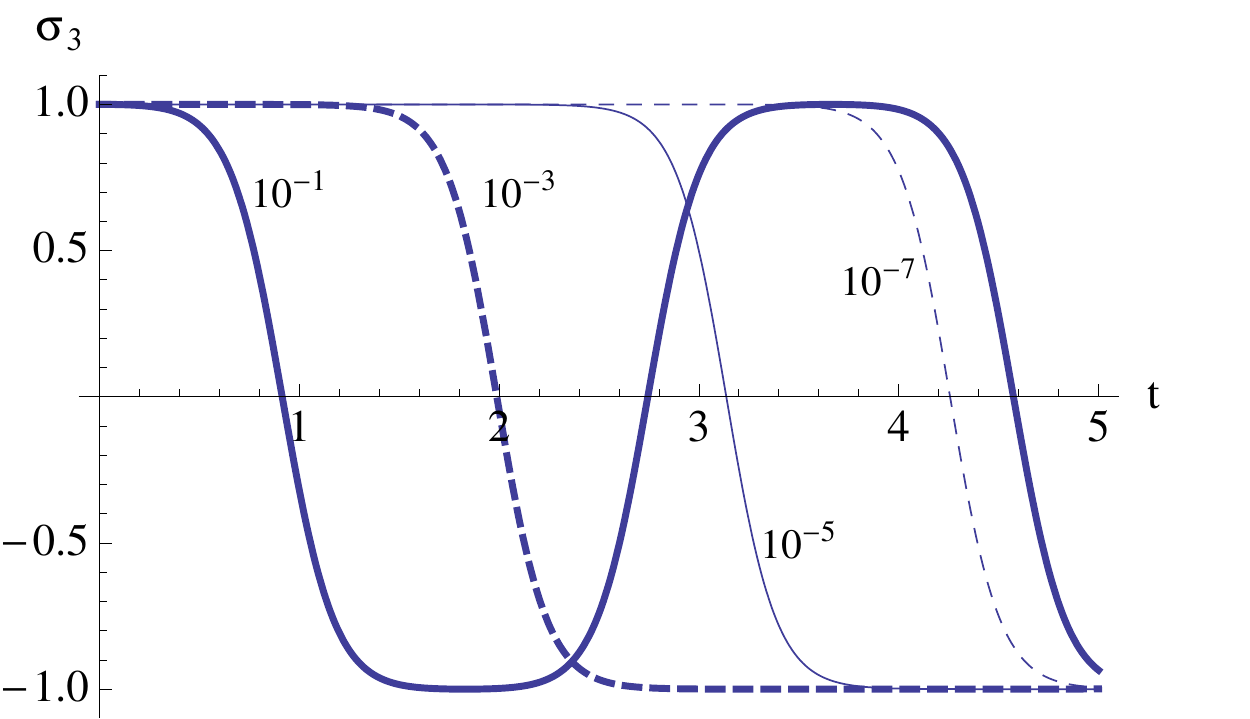} 
\caption{ The expectation of the circular polarization $\langle \sigma_3 \rangle$ of one beam versus time (in units $[n_\gamma R\,]^{-1}$) for series of small angles,
$\theta$, which measure the deviation from head-on incidence,  $1-\cos \theta=10^{-1 },10^{-3}, 10^{ -5},10^{-7}$ , with the equal spacing of the intercepts indicating the logarithmic dependence noted in text. }
 \label{fig. 2}
\end{figure}
Going back to the equations
(\ref{eom1}), (\ref{eom2}), and their counterparts for $\dot \tau_+$ and  $\dot \tau_3$, and treating them as equations for mean fields, we consider the case in which initially one beam is right-circularly polarized and the other left-circularly polarized, and in which the beams are nearly in the same direction. This  corresponds to initial values $\langle\sigma_3\rangle=1$, $\langle\tau_3\rangle=-1$ and $ \langle\sigma_+\rangle=\langle\tau_+\rangle=0$ . In fig 2 we show evolution of $\langle\sigma_3\rangle$ at later times, for a series of small angles $\theta$,
where $\theta$ is the angle between ${\bf q_1}$ and ${\bf q_2}$. When $\theta=0$, there is no polarization exchange at the mean field level.

For small angles, $\theta$ we see that there are complete helicity turnovers at regular intervals. The nonlinear oscillation is perfectly periodic, with the period as well as the time of the first crossing of the $\langle \sigma_3 \rangle=0$ axis
proportional to $-\log (1-\cos\theta)$.
For larger values of $\theta$ the time scale for the reversals is of the order $[R n_\gamma]^{-1}$, as one might have expected. But the fact that the period only expands as $\log \theta$, as $\theta\rightarrow 0$, rather than as a power is surprising.  The key to this behavior is the fact that if we go back to the equations for the case $\theta=0$, where  nothing changes in time, and linearize around this solution, we then find instabilities for exponential growth of perturbations in time.

The characteristic interaction length , L,  for polarization exchanges 
 is given by, 
\begin {eqnarray}
\,L^{-1}\approx n_\gamma R= 1.8 \times10^{-7}\Bigr [{ \sqrt{ \omega_1  \omega_2 }\over 1~{\rm  eV} } \Bigr ] 
\nonumber\\
\times \Bigr [{\rho\over {\rm 1 \,g \,cm^{-3}}} 
 \Bigr ] \Bigr [{ \sqrt{I_1\, I_2} \over 1{\rm  Watt/cm}^2 }\Bigr ] {\rm cm^{-1}}\,.
 \label{rate}
 \end{eqnarray}

\section{3. Beyond the mean-field approximation }
We address this issue by reverting to our earlier formulation with $N$ photons in a box.
We note a simple example in which the effective interaction is just $\lambda\sum \sigma_i \tau_i$ and the initial condition is $\sigma_3=1,\tau_3=-1$ 
(eigenstates).  Here nothing whatever happens at the mean-field level. The complete quantum problem for large $N$ was solved by Friedland and Lunardini \cite{fl}, who indeed find significant transitions in a time $\lambda^{-1}N$, but nothing faster. However, in the case of our slightly more complex interaction, results are quite different.
\footnote{
 The model came from the study of neutrino-neutrino interactions. The analogue to the photon polarization of our present problem is neutrino flavor. Their negative result is applicable only in cases in which there are no flavor correlations with the directions of neutrino motion.  When the latter is brought into play, the results, similar to those
 shown in fig. 3, 
may be important for neutrino flavor transmutation in the supernova core.  In consequence, the subject has generated a large number of publications, e.g. refs. \cite{rfs2}, \cite{hogwash}. The foundations for the formalism that has been used in this literature is developed, e.g., in \cite{rs}, following a rather different route than ours. }

We look at the question numerically for the case in which $H_{\rm eff}$ is given by (\ref{ha3circ}) in the $N_1=N_2=N$ case (in a box). We proceed directly from the Hamiltonian, calculating $\exp[-i H t]$ directly from the Schrodinger equation, in effect, rather than from the equations of motion
(\ref{eom1})- (\ref{eom2a}). We rescale with a time variable $t'=R^{-1} N^{-1} t$, the factor of $N$
here entering because we are reverting back to the $N$ particles-in-a-box picture, but at the same time need to readjust the volume every time we change $N$, in order to keep the densities constant. Beginning with the $\cos \theta=1$ case we find that we are able to go up to $N=1600$
effective spins in each beam, all interacting with each other.
In fig. 3 are shown computational results for $\langle \sigma_3\rangle$  for different values of $N$ for the case of initial conditions
$\langle \sigma_3\rangle=-\langle\tau_3\rangle=1$.
In the mean-field approximation
$\langle \sigma_3(t)\rangle$ would be constant in time, so the pictured behavior is \underline{all} from the deviation from the mean field value. From the equally spaced intercepts, as we increase $N$ by successive factors of two, we see that the development time is quite logarithmic in $N$ over the range between 100 and 1600. Also we see how, as $N$ becomes much larger, there will be a fairly 
well defined ``quantum break time", in the sense that the holding times at values $\sigma_3=\pm 1$ are much greater than the transition times between the two.

\begin{figure}[h] 
 \centering
\includegraphics[width=3 in]{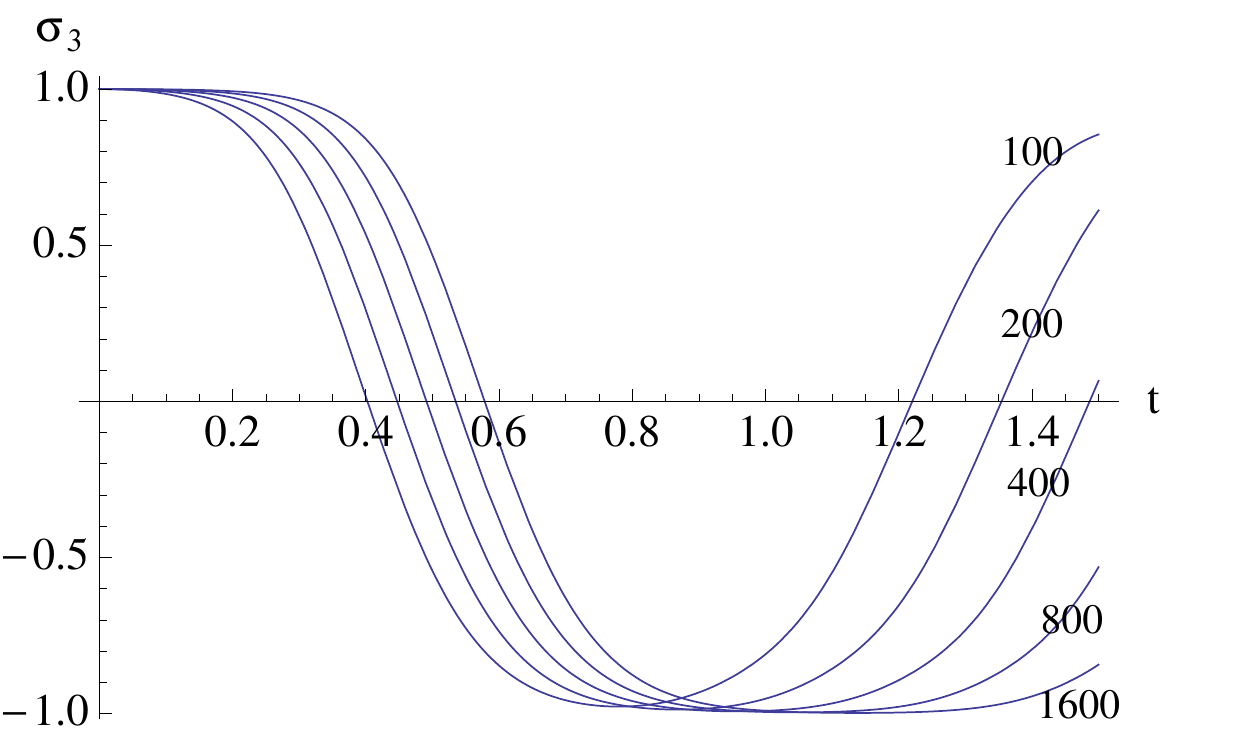} 
 \caption{ \small  For the case, $\theta=0$, the expectation of the average circular polarization per photon $\langle\sigma_3\rangle$, in one beam, plotted against time measured in units $R^{-1}$. Initially both beams are completely right-circularly polarized (i.e. with opposed spin). When$\langle\sigma_3\rangle=-1$, both beams are left-circularly polarized. The number of photons in each beam appears on the plot. The equal spacings of the intercepts with the $\langle\sigma_3\rangle=0$
axis indicates a transition time that increases as $\log N$.}   
   \label{fig. 3}
\end{figure}

This logarithmic behavior stems from beginning with a point of classical unstable equilibrium, just as in the case of the $\log (1-\cos \theta )$ behavior
within MFT noted in the last section. Examples that have the generic behavior in which the ``quantum break time" is proportional 
to the logarithm of the number of degrees of freedom are found elsewhere in the literature, for example
in models of a two-mode boson condensate in the vicinity of a dynamical instability \cite{anglin}. But we know of no proof that it cannot be faster, like
$\log[\log N]$ for example, in more complex models. Indeed, we have some evidence that it can.

In fig. 4, using the same solutions for the wave functions, we plot the quantity  $\zeta=\langle \sigma_3\tau_3 \rangle-
\langle\sigma_3\rangle\langle \tau_3 \rangle$. The plot shows directly the defect in the mean-field assumption, which requires that $\zeta=0$ at all times.

\begin{figure}[h] 
 \centering
\includegraphics[width=3 in]{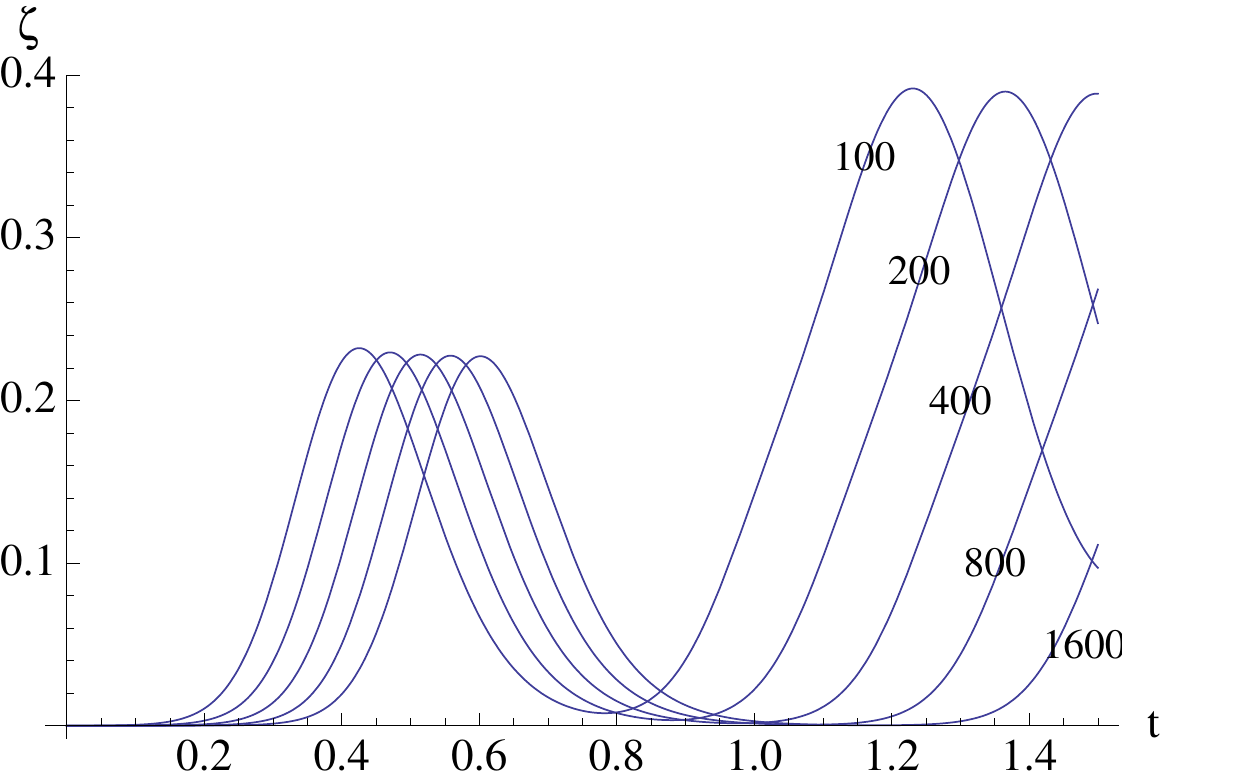} 
 \caption{ \small For the same cases used in fig. 3,  plots of  $\langle \sigma_3\tau_3 \rangle-
\langle\sigma_3\rangle \langle \tau_3 \rangle$ as a function of time and for different values of N.
}
   \label{fig. 4}
\end{figure}

The non-vanishing of $\zeta$ also indicates the development of macroscopic, $N$ body, entanglement between the two beams. Shifting terminology from photon helicity to photon spin and characterizing the initial state which has spin
projection $N$ in the ${\bf k_1}$ direction for the first beam and spin $-N$ in this same direction for the second beam, we write the initial state as $|N,-N \rangle$. At the point of complete turnover, when $ \sigma_3=-1$ the state has become $|-N,N \rangle$. Now if we compare fig. 3 with fig. 4 we see that the peaks in the parameter $\zeta$ coincide almost exactly with the points at which $\langle\sigma_3\rangle$ and$\langle\tau_3\rangle$ vanish. If at this point we had found 
$\zeta=1$, then it would tell us that the state at this time was exactly
$[\,|N,-N \rangle+|-N,N \rangle\,]/\sqrt{2}$, the classic (dead cat)-(live cat)  superposition The second peaks for the various cases shown in fig. 4 come close to the value $\zeta=.4$. 

Perhaps more interesting, we can calculate the ``entropy of entanglement" , a popular measure of entanglement \cite{entropy1} \cite{entropy2}.
We begin by tracing over one subsystemÕs coordinates, say those of beam \#2, in the density matrix for the system,
defining a reduced density matrix, $\rho_1$. Then we calculate the von-Neumann entropy corresponding 
to $\rho_1$,

\begin {eqnarray}
S_{\rm ent}= -Tr[ \rho_1 \log \rho_1]\,.
\end{eqnarray} 
In fig. 5 we plot the results for $S_{\rm ent}(t) [\log N]^{-1}$ for the same cases used in fig. 4. An interesting feature is that while the peaks,
as expected, occur at the same values of time as do the corresponding peaks of $\zeta$, the peaks of $S_{\rm ent}$ are of equal heights, while
those of $\zeta$ are of progressively increasing heights. 

Which is the better measure of the states' possible utility as an information
bearer? The author is not qualified to address this question. But there are attributes of either measure that lend themselves
to misinterpretation: In the case of $S_{\rm ent}$, we note that in our ``cat" state as defined in the introduction, a simple superposition of the
two extreme states, we have $S_{\rm ent}=\log 2$ for all values of $N$, whereas in our present calculation the peak values of $S_{\rm ent}$
increase as $\log N$.  Yet the former state is the one that we think of, loosely, as being maximally entangled.

As for the $\zeta$ measure, our interpretation of its relation to entanglement is correct only because we are in a pure state. In a mixed state, as is totally familiar in calculations
in finite temperature systems, where $\zeta$ is an ordinary correlation function, we can encounter values of $\zeta$ 
in the same ranges, but with no quantum entanglement.

\begin{figure}[h] 
 \centering
\includegraphics[width=3 in]{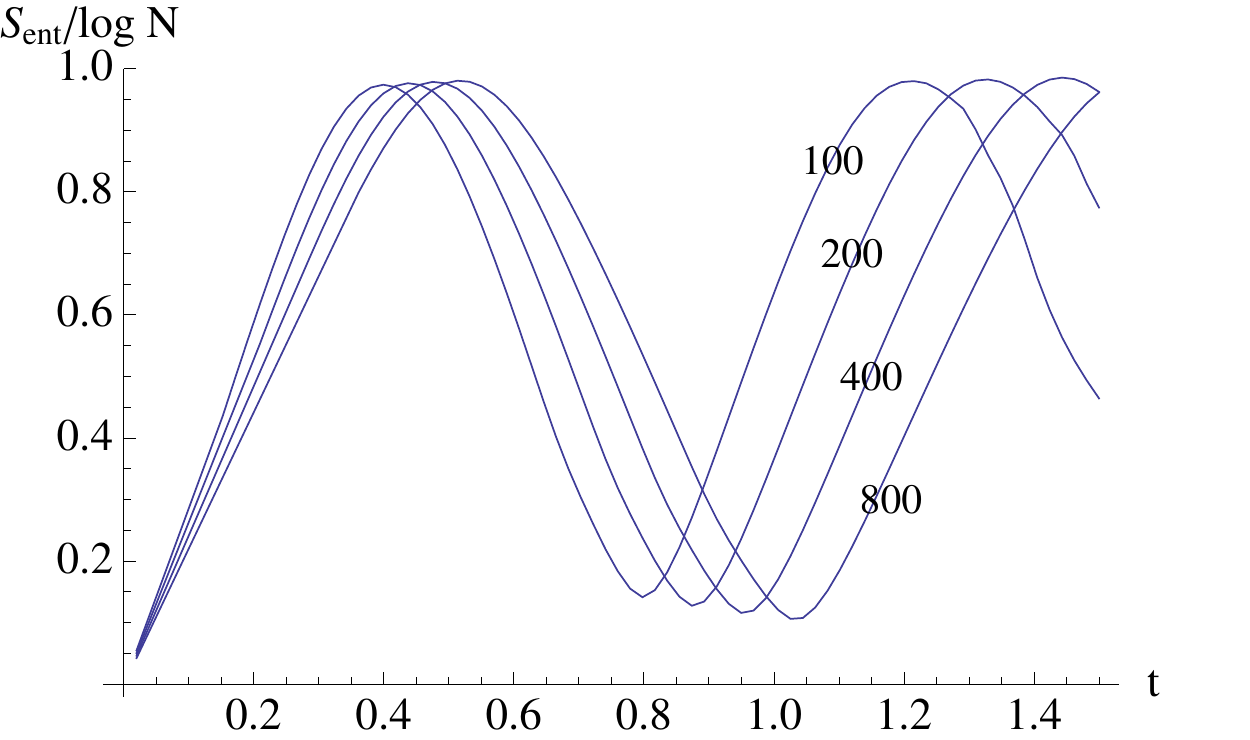} 
 \caption{$ S_{\rm ent} [\log N]^{-1}$  as function of time and for different values of $N$.
}
   \label{fig. 5}
\end{figure}

Having looked at both mean-field solutions and exact solutions, but the latter only for $N\le 1600$, we can ask if there is a viable approximation scheme that
can recapture the quantum effects and that is not subject to the $N$ limitation.
A possible approach, which was applied to the $\theta=0$ case of the model of this section \footnote{This previous work addressed polarization exchange in photon-photon interactions, but the interaction used was the Heisenberg-Euler effective interaction in vacuum \cite{iz}, rather than an atomic gas. See also the work of Kotkin and Serbo \cite{ks}. When, and only
when, $\theta=0$, $H_{\rm eff}$ has the same polarization dependence as found for the vacuum case.  The last section of ref. \cite{rfs1} addressed the logarithmic time scale for
polarization exchange in this case.}, 
is through iteration of the equations of motion to get a hierarchy of equations, still linear in ${d/ d t}$, in which the right hand sides are now expectations of higher order polynomials in the $\sigma_i$ and $\tau_i$
variables \cite{rfs1}. 
A plausible and systematic procedure was applied for factorizing these equations into the products of expectation values of linear and quadratic forms in the density matrix, producing
a closed set of equations, where one term, from an evaluated commutator, bears a $N^{-1}$ factor. 
For the initial conditions ($\theta=0$) corresponding to those used in fig. 3., the solution to these equations gave a good fit to the plots
of fig. 3  for the quantity  $\langle \sigma_3 \rangle$, indicating some correct account of quantum effects.  But it gives zero for $\zeta$, rather than the entanglement shown in fig. 4.

\section{4. More quantum examples.}

We have also calculated the evolution for the case of small angle $\theta$ deviations from head-on and the same initial conditions, in the hopes of comparing with, say, the mean-field data shown in fig. 2. Here we reach our computational limit at $N_\gamma=128$ in each beam (much reduced from the head-on case because the operator $\sigma_3+\tau_3$ is no longer conserved, leading to a much bigger set of $N^2$ states that enter the calculation).

Because of the limitations on $N$ we are less certain of the meaning of the results in this case.  They are at least consistent with the following picture: at a fixed very small angle the deviations from mean-field will dominate in some region of smaller $N$. In a region of transitional $N$ these deviations will give great distortions in the shape and completeness of the turnover, and for $N$ very large 
they will not enter at all until we get to times of order $\log N$. By this time, even for very small values of $\theta$, the polarization oscillations will already have gone through many cycles, according to the mean-field calculations in sec.3. 
To put the problem more dramatically, suppose we actually tried to produce the nice behavior shown in fig. 4 in a laboratory, and we used, say, 10$^{21}$
photons at any moment in a reaction cylinder of length 1 cm. and radius 1 mm. Then it appears that we might have to start with $1-\cos \theta=0$ to within one part in $10^{21}$, roughly, to avoid being obscured by dominant mean-field effects.

As an example of how to overcome this absurdity, and more generally to demonstrate the variety of behaviors that these models can produce, we show
explicitly the results of one more variant. We go back to the form of the interaction ( \ref{ha2}) where now the $\sigma_i$'s and $\tau_i$'s are 
constructed with respect to our original plane-polarized basis, and we begin with eigenstates $\sigma_3=1$ and $\tau_3=-1$. 
We are then in MFT equilibrium for \underline{any} value of $\theta$ and the equilibrium is unstable at in a range near $\theta=0$, so that the quantum effects grow rapidly. But now we analyze the system at a later time by computing the expectations of
 $\sigma_3'$ and $\tau_3'$, as referred to polarization bases rotated, with respect to the choice used in the preparation of the states, 45$^\circ$ around the respective ${\bf q}$ and ${\bf q'}$ directions. In fig. 6 we show some plots of $\zeta'$, \begin{eqnarray}
\zeta' =\langle \sigma_3' \tau_3' \rangle-\langle \sigma_3'\rangle\langle \tau_3' \rangle
\end{eqnarray}
for $\cos \theta=.96$, showing a rapid rise in  $\zeta'$ in time, for different values of N.
Fig. 7 shows behavior over a longer time and for $\cos\theta=1$.
We see a rapid rise at the beginning, followed by a plateau with $\zeta'$ almost exactly $.5$.
The initial rise time here is logarithmic in $N$, as expected. The length of the plateau, however, unexpectedly increases linearly with $N$. 
In the plateau region we have$\langle \sigma_3' \rangle\approx \langle \tau_3' \rangle\approx 0$.
\begin{figure}[h] 
 \centering
\includegraphics[width=3 in]{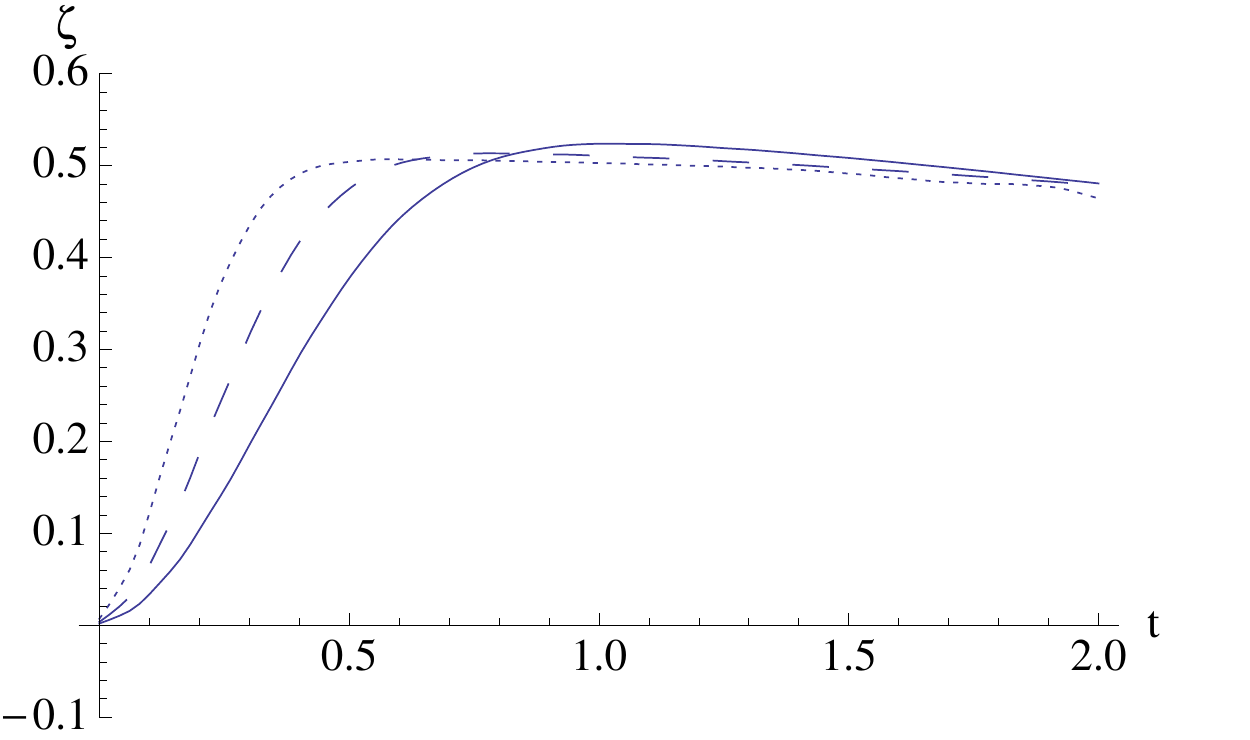} 
 \caption{ \small For $\cos \theta=.96$, plots of the entanglement measure $\zeta'$ against scaled time. Dotted curve is for N=15; dashed for N=30; solid for N=60 }
   \label{fig. 5}
\end{figure}

\begin{figure}[h] 
 \centering
\includegraphics[width=3 in]{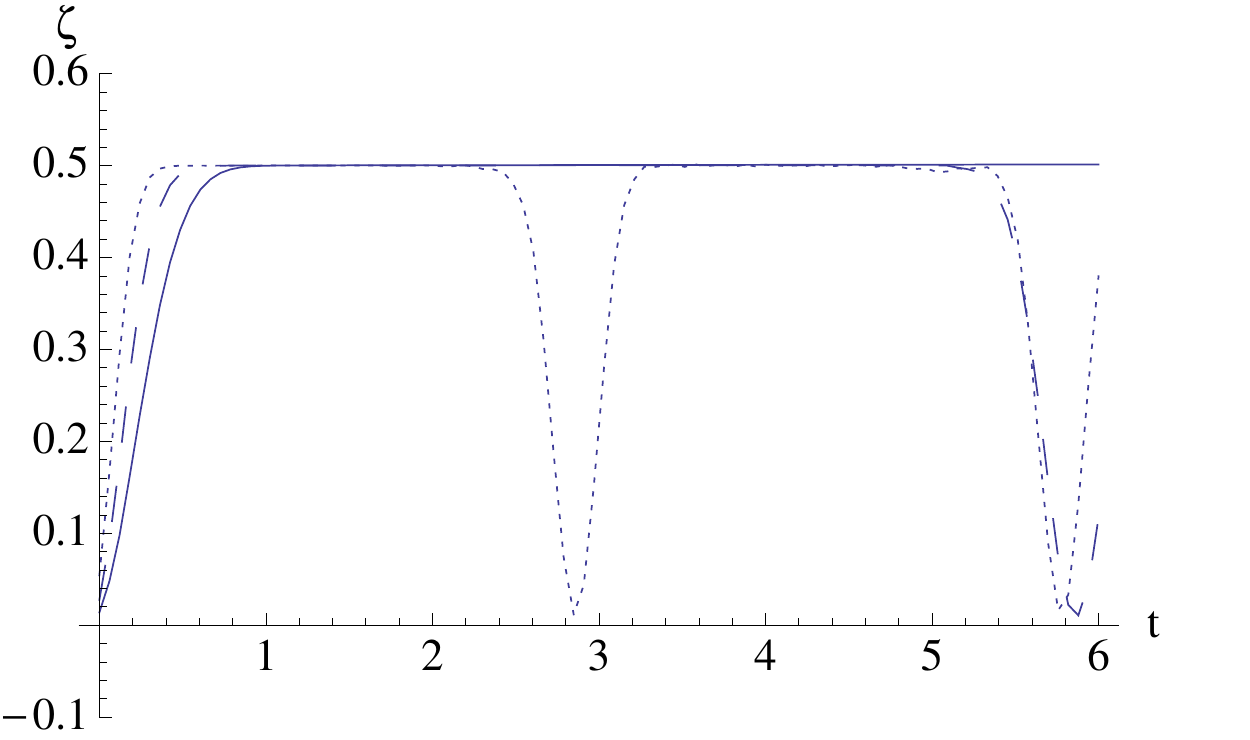} 
 \caption{ \small Same as in fig. 6, except for $\cos \theta=1$, and plotted over 3 times the time interval. The rise time to get to $\zeta'\approx .5$
 goes approximately as $\log N$, as expected. Unexpected is the fact that, as $N$ increases, the ``hang-time", or the the period in which $\zeta'$
 spends at this level before making a short round trip to zero and back, grows linearly with $N$.  }
   \label{fig. 6}
\end{figure}

A value of $\zeta'=1$, in  a pure state, would have signaled the simple ``cat" state, $| \psi \rangle$, referred to our original basis,
\begin{eqnarray} 
| \psi \rangle=2^{-1/2}(\,| \uparrow,  \uparrow\rangle+ | \downarrow,  \downarrow\rangle\,)
\end{eqnarray}

We have been able to get a value as high as $\zeta'=.7$ for very short periods of time by changing initial conditions in the above by a little, but we
are not able to explore the parameter space systematically to see if $\zeta'=1$ is achievable. 

In further numerical experiments, choosing some initial conditions that are not quite ones of MFT equilibrium, and that also include some elliptical polarization, we get reasonably large values of $\zeta$ in times that appear to be are either finite as $N\rightarrow \infty$ or growing
more slowly than $\log N$. Because now the initial state is not in an MFT equilibrium state this latter turnover occurs in time that is nearly independent of $N$.   But these results are more irregular-looking than those shown in the figures of this section and the last, and obtaining them involved the solution of 
 $N^2$ coupled non-linear differential equations (as did the calculations for figs. 5 and 6), rather than the $2N $ needed for the solutions in the last section.

\section{5. Pulse behavior}
One of the drastic idealizations in all of the above was in taking a system with plane waves in a box (with implied periodic boundary conditions), and turning on the interaction among all particles at time zero. In our desired experiment we would have to get them in and out, and different photons come and leave at different times. 

Take the beams to be very nearly co-linear, $(1-\cos\theta)<<1$ in (\ref{eom1})-(\ref{eom2a}), and (nearly) totally overlapping over the region $0<z<L$. We have no insight into how easy, hard, or impossible the latter would be to achieve in a laboratory. But in any case, we have assumed that the idealization is adequate to predict the results of an achievable 
case in which the interaction region is that of the intersection of two narrow cylinders of width $d$ at a relative angle 
$\theta$ such that the intersection length is of order $L\sim d/\theta$. 

For clarification we pose a time and space dependent problem in which at $z=0$ we take superposed beams of opposite helicity, both moving nearly in the +z direction, and entering the reaction region just before the initial time in the calculation. The governing equations for $\vec \sigma(z,t)$, $\vec \tau(z,t)$  are still (\ref{eom1}) - (\ref{eom2a}) but with 
$\partial/\partial t$ replaced by $\partial/\partial t+\partial/\partial z$. The boundary conditions are the values $\sigma_3(0,t)$, which rises rapidly to unity and then is maintained at that value, and $\sigma_+(0,t)=0$. The $\tau$ values are the exact negatives of the $\sigma$'s. The initial condition has a short region in which a smoothed leading edge of the
beam has entered the very beginning of the interaction space but everything beyond is zero. In fig. 5 we show
the results. The polarization very rapidly evolves into a standing wave pattern. \footnote{ We emphasize that the photons themselves are not standing waves, they run in the z direction indefinitely; it is just the polarizations that have the standing pattern}

Thus in our MF case, of sec. 3, if we choose the correct length for the interaction region, and send in a pulse that is in the product polarization state specified above, 
we can expect to get out a pulse in which there has been total polarization exchange between the two superimposed beams. 

We expect similar standing wave behavior in the quantum cases, i.e. in a beyond-the-mean-field calculation.  But we do not have the power to do the quantum calculation with combined space and time dependence in order to verify this expectation.

\begin{figure}[h] 
 \centering
\includegraphics[width=3 in]{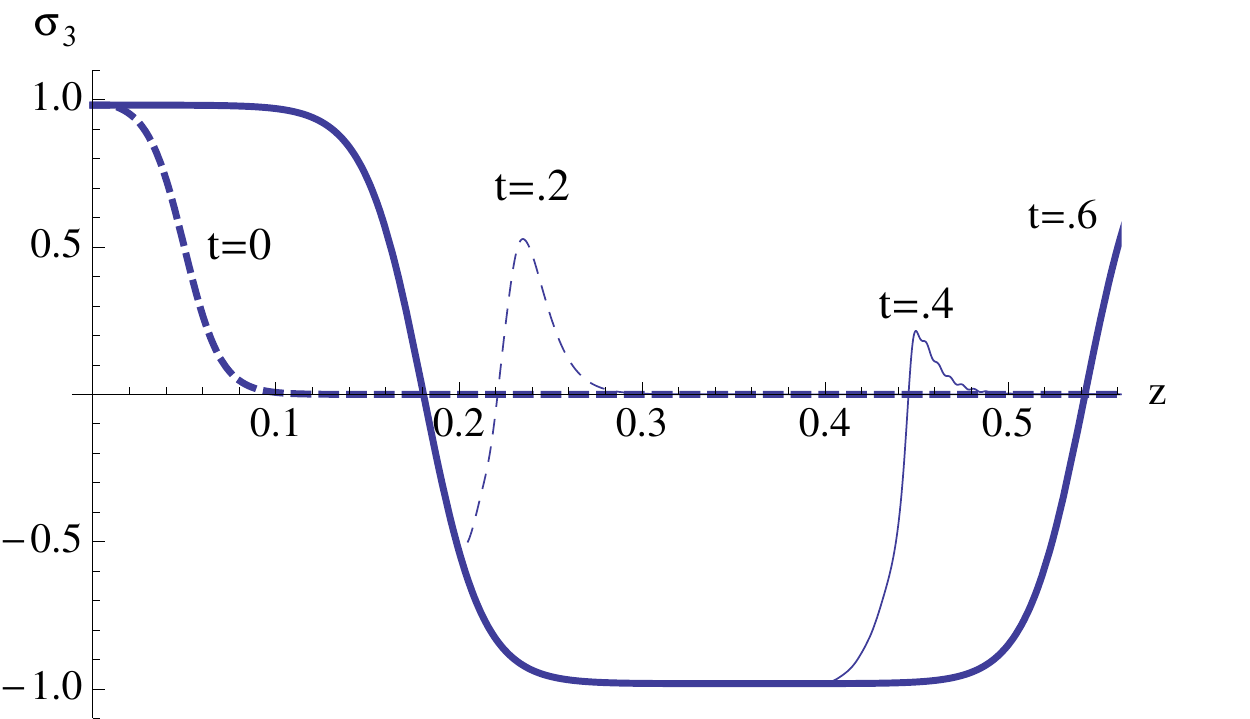} 
 \caption{ \small For $\cos \theta=.99$ plots of  $\langle \sigma_3 \rangle $ as a function of the distance $z$ from the injection point at $z=0$, for a series of times after the beams, both coming from the left, are turned on. The heavy dashed curve at near distances is the (arbitrary) assumed leading profile of the beam at its entry shortly before $t=0$, it has not at this point been in the medium long enough to have have changed polarizations appreciably. The profiles at later times show the effects of interaction. Results for $t\,>.6$ are indistinguishable from the heavy, $t=.6$, profile. Note that the standng pattern,
region by region, develops at times only slightly greater than the light travel time $z$. The profiles for $\tau_3$
are the exact negatives of the values shown. }
   \label{fig. 7}
\end{figure}
 
 As mentioned previously, beams meeting head-on at a very small angles, $\theta=\pi-\alpha$, where $\alpha<<1$ show
 behavior identical that of the $\theta<<1$ case, in the purely time dependent evolution in a periodic box. But 
 when we instead let them enter at time zero from opposite ends into a finite interaction region, we suddenly have a
 formidable computational problem even at the mean-field level, due to the mixed boundary condition; prescribed values for at $z=0$ for the right-moving beam and at $z=L$ for the left-moving beam. Suffice to say that we have made no progress toward a 
solution, possibly because the instability of the $\theta=\pi$ equations now manifests itself through very sensitive
dependence on boundary conditions. 

However we have shown how it is possible in some cases to  proceed from studying the time development of two interacting waves confined to a box, with
periodic boundary conditions, to studying the space-time dependence of the beams after their introduction on one side of a region containing the transforming medium. The pure nonlinear oscillations in time in the first instance were transformed into into standing waves in space in the second instance.
\section{6. Discussion}

In principle, photon-photon interactions in a medium can produce polarization-entangled beams, leaving the medium untouched. There 
appears to be much flexibility in the encoding and processing of information in the interactions of such beams, by virtue of the considerable landscape of polarization 
phenomena that are available and of the ease of manipulation of polarization parameters for a single beam.  

We have not found many close analogues to our system in the very large literature on non-linear photon effects and systems of entangled photons.
Of early works that deal with the entanglement of photons with photons we mention that of ref. \cite{ima}. But the system discussed in this paper is quite 
different from that of ours; it has subsidiary applied fields as part of the mechanism, uses tuning vey near resonances as an integral part
of the procedure, and is not specifically concerned with polarizations.

We do mention an interesting realization in the laboratory \cite{polzik} of a state that has some elements in common with the ones that
we produce theoretically. In brief, we describe a result of this work: the beginning state is a big set of N up-spins that
is subjected to radiation which (some of the time) turns just one into a down-spin. The excitation is coherently spread over the different spins.
Then more fields are applied that mix the up-states with a 50-50 superposition of up and down,
with the down state going into the orthogonal combination. Next, in repeated trials, the distribution of the differences of the up and down numbers is measured,
and its variance is proportional to $N^{1/2}$, as compared with unity for that produced by a simple simple spin flip when the subsequent
manipulation is omitted, a clear quantum effect. 
 
In this case it is not useful to try to  describe the quantum state as ``entangled", a term that is defined only with respect to a division of the degrees of freedom into two sets (as in our two beams). But it leads us to calculate one more measure of the quantum nature of the outcome of the
calculation described in sec. 3; the variance of ($N_\uparrow-N_\downarrow $) for just one of the beams.  We find peaks at almost exactly the positions shown in
figs. 3 and 4, with the curves beginning at zero at $t=0$, and the maxima growing as $N$.

Interesting as we find the different behaviors sketched in this paper, the issue of practicality has not been settled here, to say the least. Because atomic hydrogen is the only medium in which we can
do a complete calculation, it is what we have used in this paper in order to provide a rough estimate of requirements. Other media
should do better, particularly for one or both of the beams tuned near a resonance frequency. 

Finally, it would be of great interest to know if there are some regions in the landscape of initial conditions in which a large entanglement
develops in a time that approaches a finite limit as $N\rightarrow \infty$ (or perhaps goes, e.g., as $\log[\log N]$ instead of increasing as $\log N$).
We have some indications that there are, but at the moment we have insufficient computing power to investigate completely. 

\section{Acknowledgement}  
I am grateful to Daniel Loss for pointing out the possible relevance of ref. 17, and to Jonathan Keeling for bringing to my
attention the works of ref 1-3.  This work was supported 
in part by NSF grant PHY-0455918. 

\end{document}